\documentclass[letterpaper,pra,aps,twocolumn,showpacs]{revtex4}
\usepackage{graphicx,bm,color}
\usepackage{mathrsfs}
\usepackage{amsmath}
\usepackage[normalem]{ulem}
\usepackage{subfigure}

\newcommand{\bfk}{{\bm k}}
\newcommand{\bfp}{{\bm p}}

\begin{document}

\title{$d_{xy}$-density wave in fermion-fermion cold atom mixtures}
\author{Chen-Yen Lai$^1$}
\author{Wen-Min Huang$^{1,2,3}$}
\email[Email: ]{wmhuang0803@gmail.com}
\author{David K. Campbell$^4$}
\author{Shan-Wen Tsai$^1$}
\affiliation{$^1$ Department of Physics and Astronomy, University of California, Riverside, CA 92521, USA}
\affiliation{$^2$ Zentrum f\"ur Optische Quantentechnologien and Institut f\"ur Laserphysik, Universit\"at Hamburg, 22761 Hamburg, Germany}
\affiliation{$^3$ Hamburg Centre for Ultrafast Imaging, Luruper Chaussee 149, Hamburg 22761, Germany}
\affiliation{$^4$ Department of Physics, Boston University, Boston, Massachusetts 02215, USA}

\begin{abstract} 
We study density wave instabilities in a doubly-degenerate Fermi-Fermi mixture with $SU(2)\times SU(2)$ symmetry on a square lattice. 
For sufficiently large on-site inter-species repulsion, when the two species of fermions are both at half-filling, two conventional ($s$-wave) number density waves are formed with a $\pi$-phase difference between them to minimize the inter-species repulsion. 
Upon moving one species away from half-filling, an unconventional density wave with $d_{xy}$-wave symmetry emerges. 
When both species are away from the vicinity of half-filling, superconducting instabilities dominate.  
We present results of a functional renormalization-group calculation that maps out the phase diagram at weak couplings. 
Also, we provide a simple explanation for the emergence of the $d_{xy}$-density wave phase based on a four-patch model. 
We find a robust and general mechanism for $d_{xy}$-density-wave formation that is related to the shape and size of the Fermi surfaces. 
The density imbalance between the two species of fermions in the vicinity of half-filling leads to phase-space discrepancy for different inter-species Umklapp couplings. 
Using a phase space argument for leading corrections in the one-loop renormalization group approach to fermions, we show that the phase-space discrepancy in our system causes opposite flows for the two leading intra-species Umklapp couplings and that this triggers the $d_{xy}$-density-wave instability.
\end{abstract}

\date{\today}

\pacs{67.85.Lm, 64.60.ae, 71.10.Hf, 71.45.Lr}

\maketitle

\section{Introduction}
Experimental realizations of quantum degenerate Fermi-Fermi mixtures have opened a new arena for the study of quantum many-body phenomena in cold atom systems~\cite{Shin,Wille:2008gq,Spiegelhalder:2009hw,Spiegelhalder:2010hq,Hara:2011gq}. 
In these multi-component systems, many different types of fermionic superfluids have been proposed and studied~\cite{Yip:2011eq,Inaba:2012jd}. 
For population imbalanced mixtures, a number of exotic unconventional pairing states, some of which do not occur in normal condensed-matter systems, have been investigated theoretically, including Fulde-Ferrell-Larkin-Ovchinnikov~\cite{Larkin:1965uw} superfluidity in the vicinity of phase separation~\cite{Radzihovsky}, interior gap superfluidity~\cite{Liu}, and $p$-wave pairing~\cite{Bulgac,Raghu,Patton,Lai}. 
However, once strong Fermi surface (FS) nesting is present, instead of superconducting pairing, density wave instabilities dominate~\cite{Lai,WM} and may exhibit higher angular momentum order.
Unconventional density waves (density waves with non-zero angular momentum)~\cite{Campbell,Marston,Nayak} have recently been proposed theoretically for ultracold fermionic atoms. 
By loading fermionic dipolar atoms~\cite{Lu} and molecules~\cite{Ni,Chotia,Wu} onto optical lattices, unconventional density-wave instabilities may arise from the long-range and anisotropic dipole-dipole~\cite{Yamaguchi,Mikelsons,Parish,Bhongale1} or quadrupolar~\cite{Bhongale3} interactions. 
By creating a Kagome optical lattice~\cite{Jo}, a bond order wave can be stabilized,resulting from sublattice interference effects~\cite{Kiesel1}, in the presence of a sufficiently large nearest-neighbor repulsive term~\cite{Wang12,Kiesel2}.

\begin{figure}[t!]
\begin{center}
\includegraphics[width=0.45\textwidth]{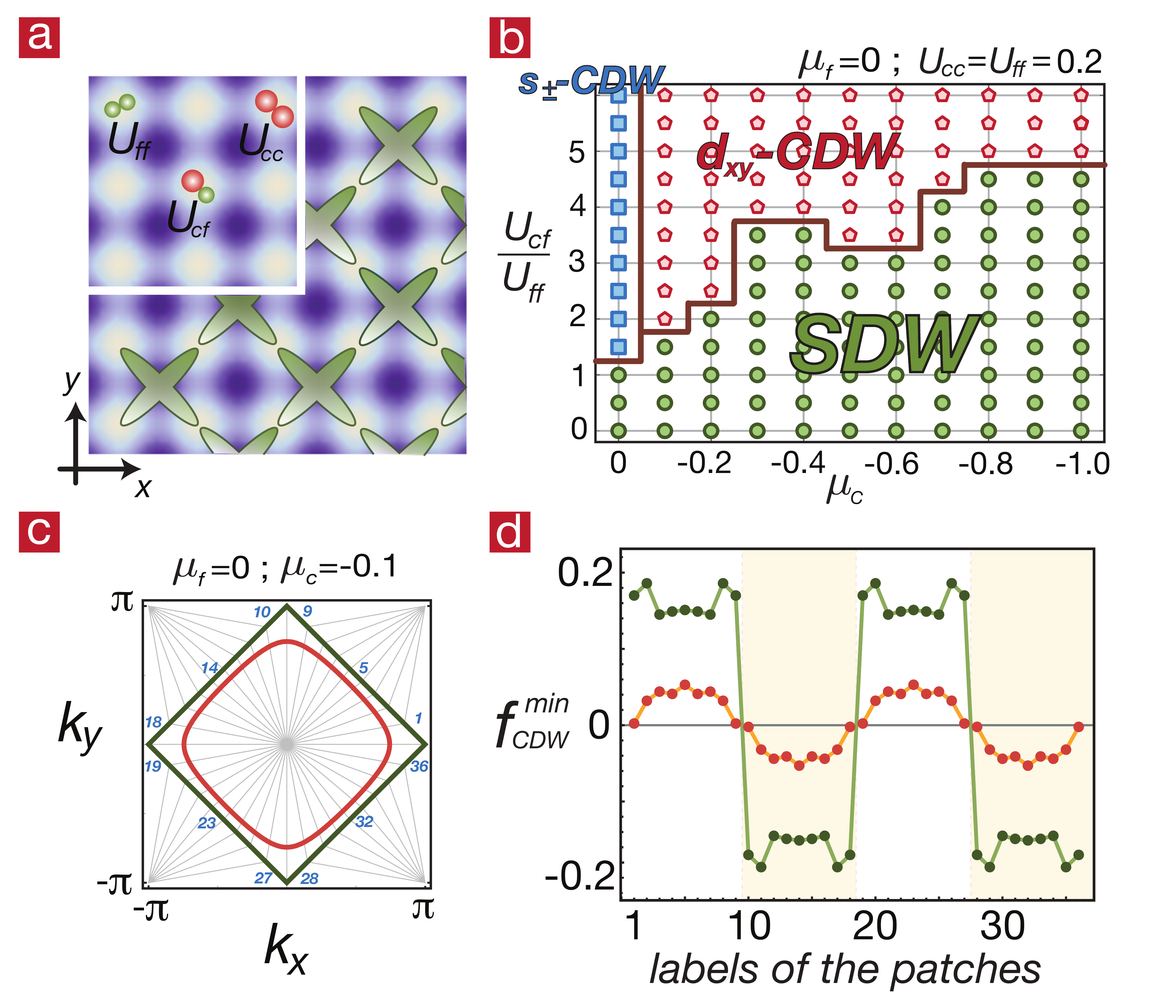}
\caption{(Color online) 
(a) Contour plot of a square optical lattice with a fermion-fermion mixture, where lighter regions correspond to lower potential energy and darker to higher energy, and real-space picture of the $d_{xy}$-CDW state for one species of fermions. 
(b) The phase diagram is parameterized by the chemical potential of $c$-fermions, $\mu_c$, and the ratio of inter- and intra-species interactions, $U_{\rm cf}/U_{\rm ff}$. 
(c) The FS patches used in this study ($M=36$ on each FS), as well as the FS for $\mu_f=0$ and  $\mu_c=-0.1$, and (d) corresponding $d_{xy}$ order parameter for $f$- and $c$-fermions (orange and green dots, respectively).
}
\label{fig:phase}
\end{center}
\end{figure}

In this paper, we study density wave instabilities of a doubly-degenerate fermion mixture, such as $^6$Li and $^{40}$K~\cite{Spiegelhalder:2010hq,Taglieber}, on a square lattice. As illustrated in the insert figure of Fig.~\ref{fig:phase}a, we describe the system as a simple Hamiltonian, $H=H_t^c+H_t^f+H_{\rm int}$, where
\begin{eqnarray}
&&\hspace{-0.4cm}H_t^a=-t\sum_{\langle ij\rangle,\alpha}\left(a^{\dag}_{i\alpha}a_{j\alpha}+{\rm H.c.}\right)-\mu_a \sum_{i}n_{ai},\label{eq:Ht}\\
&&\hspace{-0.4cm} H_{\rm int}=\sum_{i}U_{\rm cc}n_{ci\uparrow}n_{ci\downarrow}+U_{\rm ff}n_{fi\uparrow}n_{fi\downarrow}+U_{\rm cf}n_{ci}n_{fi}\label{eq:Hint},
\end{eqnarray}
where $a=c/f$ stands for the $c/f$-fermions of the two species, $\alpha$ is the spin index. $H_t^a$ is described the tight-binding model with a nearest-neighbor hopping amplitude $t_c=t_f=t$, and $\mu_a$ is the chemical potential of the $a$-fermions. We also include the inter- and intra-species interactions, represented as $U_{\rm cc(ff)}$ and $U_{\rm cf}$ with the definition $n_{ai}=n_{ai\uparrow}+n_{ai\downarrow}=\sum_{\alpha}a^{\dag}_{i\alpha}a_{i\alpha}$.

%$\langle ij\rangle$ represents nearest-neighbor pairs of sites, $n_{ai}=n_{ai\uparrow}+n_{ai\downarrow}=\sum_{\alpha}a^{\dag}_{i\alpha}a_{i\alpha}$ and $\mu_a$ is the chemical potential of the $a$-fermions. 
%For simplicity, we set nearest neighbor tunneling $t_c=t_f=t$. 
 
We consider the limit of weak repulsive interactions and employ a standard  functional renormalization-group (FRG) method~\cite{Shankar:1994vy,Metzner:2012jv,Zanchi:2000vv,Tsai:2001bh}, which allows us to treat the inter- and intra-species interactions on an equal footing. 
We find that for large intra-species repulsion, the behavior is reminiscent of a single $SU(2)$ species, i.e. two independent spin-density waves (SDW), while both species are near half-filling. 
However, in the presence of an inter-species repulsion sufficiently larger than the intra-species repulsion, two unconventional charge-density waves (CDW) start to emerge, and the phase diagram is showed in Fig.~\ref{fig:phase}b. 
When both species are at half-filling, the ground state consists of two conventional $s$-wave number density waves, here denoted as charge-density-wave (CDW) in analogy with number density waves in electron systems. 
As one species is moved away from half-filling, a $d_{xy}$-wave CDW instability becomes dominant. 

To further elucidate the physics behind the $d_{xy}$-CDW formation, we develop a simple four-patch model with intra- and inter-species Umklapp couplings that provides an analytical understanding of the competition between $s$- and $d_{xy}$-CDWs. A density imbalance between the two species of fermions in the vicinity of half-filling leads to different effective phase spaces for the different inter-species Umklapp couplings. Using reasoning similar to Shankar's phase space argument for determining one-loop RG flow equations~\cite{Shankar:1994vy}, we explain how the phase-space discrepancy for the inter-species Umklapp couplings leads to opposite RG flows for the different intra-species Umklapp couplings, triggering the $d_{xy}$-CDW instability. 
By varying the densities of the two species around half-filling, we also study the range of fillings for which this regime of unconventional  $d_{xy}$-CDW can be realized. 

The remainder of the paper is arranged as follows. 
In Sec.~\ref{sct:modelandfrg}, we introduce our theoretical approach, the functional renormalization group (FRG) method. 
In Sec.~\ref{sct:result}, the main part of this paper, we present a detailed discussion of the phase diagram of the system and physical mechanism leading to the  $d_{xy}$-CDW instability. 
In Sec.~\ref{sct:exp}, we discuss a proposed experimental realization of this model and the detection of $d_{xy}$-CDW order parameter. 
Finally, in Sec.~\ref{sct:conclusion} we present a summary of our present results and discuss prospects for future work.

%\section{Model and Formalism}\label{sct:modelandfrg}

%\subsection{Hamiltonian}
%We start with a simple Hamiltonian describing a fermion-fermion mixture on a square lattice, $H=H_t^c+H_t^f+H_{\rm int}$, where
%\begin{eqnarray}
%&&\hspace{-0.4cm}H_t^a=-t\sum_{\langle ij\rangle,\alpha}\left(a^{\dag}_{i\alpha}a_{j\alpha}+{\rm H.c.}\right)-\mu_a \sum_{i}n_{ai},\label{eq:Ht}\\
%&&\hspace{-0.4cm} H_{\rm int}=\sum_{i}U_{\rm cc}n_{ci\uparrow}n_{ci\downarrow}+U_{\rm ff}n_{fi\uparrow}n_{fi\downarrow}+U_{\rm cf}n_{ci}n_{fi}\label{eq:Hint},
%\end{eqnarray}
%where $a=c/f$ stands for the $c/f$-fermions, $\alpha$ is the spin index, $\langle ij\rangle$ represents nearest-neighbor pairs of sites, $n_{ai}=n_{ai\uparrow}+n_{ai\downarrow}=\sum_{\alpha}a^{\dag}_{i\alpha}a_{i\alpha}$ and $\mu_a$ is the chemical potential of the $a$-fermions. 
%For simplicity, we set nearest neighbor tunneling $t_c=t_f=t$. 

\begin{figure}[t]
\begin{center}
\includegraphics[width=0.45\textwidth]{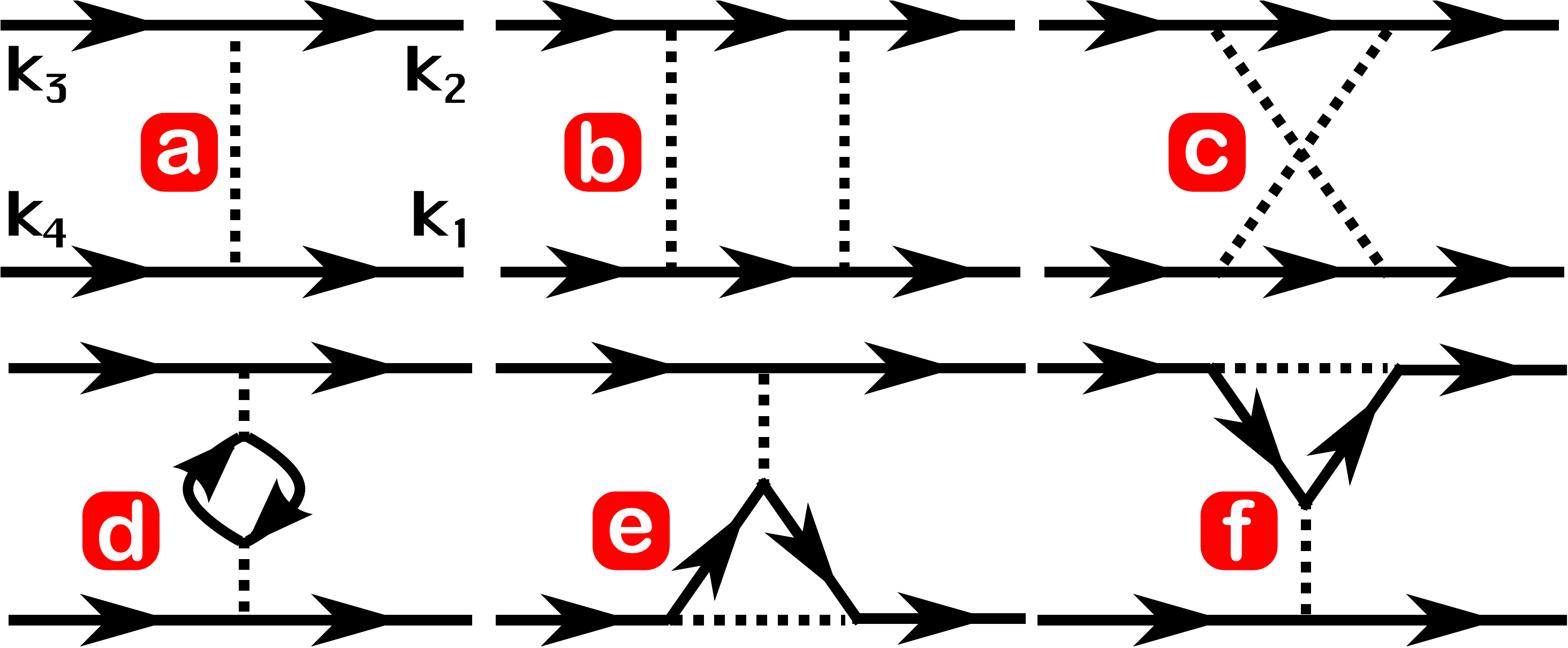}
\caption{(Color online) (a) A four-point vertex function $g(\bfk_1,\bfk_2,\bfk_3)$. (b)-(f) One-loop diagrams that contribute to the renormalization of a four-point vertex.}
\label{fig:rgeq}
\end{center}
\end{figure}

\section{The Functional Renormalization Group for two species of fermions}\label{sct:modelandfrg}
We consider the limit of weak repulsive interactions, $0<U_{\rm cc}, U_{\rm ff}, U_{\rm cf}< \mathcal{W}$, where $\mathcal{W}$ is the full energy band width. We employ a standard FRG method~\cite{Shankar:1994vy,Metzner:2012jv,Zanchi:2000vv,Tsai:2001bh} to obtain the low-energy behavior for this system.
We calculate the corrections to the interaction vertices to one-loop\cite{Zanchi:2000vv}, which correspond to the diagrams shown in  Figs.~\ref{fig:rgeq}(b)-\ref{fig:rgeq}(f). 
For the fermion-fermion mixture we consider, each solid line in these diagrams also carries a spin ($\alpha = \uparrow, \downarrow$) and a species ($a=c, f$) index. 
Since the Hamiltonian, Eqs.~(\ref{eq:Ht}) and~(\ref{eq:Hint}), has $SU(2)$ symmetry for each species, it is enough to calculate only vertices in which fermions with antiparallel spins interact, as vertices involving fermions with the same spin can be obtained from these by imposing $SU(2)$ symmetry. 
Depending on the species indices, we have intra-species interaction vertices for each species or inter-species interaction vertices. 
Since the Hamiltonian also has global $U(1)$ symmetry and interactions are of the density-density type, to each fermion annihilated (incoming fermion line) corresponds a fermion with the same spin and species indices being created (outgoing line). 
Thus we can write any effective interaction term in the form:  $g_{abba}({\bm k}_1,{\bm k}_2,{\bm k}_3,\ell)\psi^{\dag}_{a\alpha}({\bm k}_1)\psi^{\dag}_{b\beta}({\bm k}_2)\psi_{b\beta}({\bm k}_3)\psi_{a\alpha}({\bm k}_1+{\bm k}_2-{\bm k}_3)$, in momentum space, with fermionic field $\psi({\bm k}=(k_x,k_y))$, and $\ell=\ln(\mathcal{W}/\Lambda)$, with $\Lambda$ the running UV energy cut-off. 
All interaction terms require momentum conservation as ${\bm k}_1 + {\bm k}_2 = {\bm k}_3 + {\bm k}_4$ shown in Fig.~\ref{fig:rgeq}(a). 
The species indices appear as a pair with the same spin; this follows from the density-density nature of the inter-species interaction in the Hamiltonian, as mentioned above.

To solve the RG flow equations numerically, we discretize the Fermi surfaces into patches~\cite{Zanchi:2000vv} and label the interaction vertices in terms of patch-indices, instead of momenta, writing $g(\bfk_1,\bfk_2,\bfk_3,\ell)\rightarrow g(i_1,i_2,i_3,\ell)$. 
In our numerical calculations, we divide each Fermi surface into 36 patches. 
The configuration of the Fermi surface patches is illustrated in Fig.~\ref{fig:phase}(c). 
After carrying out the one-loop RG calculations, we obtain the flow of four-point vertices in terms of the RG scale, $\ell$, and three independent momenta, $g(\bfk_1,\bfk_2,\bfk_3,\ell)$. 
The momenta are discretized in the numerical implementation, but we describe here the general analysis in terms of momenta. 
The signature for the occurrence of an instability of the Fermi liquid state in the RG approach is the development of run-away flows. 
To determine the dominant instability, we look at the quartic interaction terms in the effective action in the form
$\sum_{\bfk,\bfp}\mathcal{V}^{(\ell)}_{\rm op}(\bfk,\bfp)\hat{\mathcal{O}}^{\dag}_{\bfk}\hat{\mathcal{O}}_{\bfp}$, with $\hat{\mathcal{O}}_{\bfk}$ 
being a term bilinear in the fermion fields and corresponding to the order parameter (OP) of superconducting pairing (SC), charge density wave (CDW), spin-density wave (SDW), ferromagnetism (FM) or Pomeranchuk (PI) instabilities\cite{Zanchi:2000vv,Zhai:2009br}. 
The amplitudes for each instability channel are matrices and are related to the couplings $g$ by
\begin{eqnarray}
\mathcal{V}^{(\ell)}_{SC_{s(t_0)}}(\bfk,\bfp)&=&\frac12\left[g(\bfk,-\bfk,-\bfp,\bfp,\ell)\pm g(-\bfk,\bfk,-\bfp,\bfp,\ell)\right]\nonumber\\
\mathcal{V}^{(\ell)}_{SDW}(\bfk,\bfp)&=&-g(\bfp,\bfk,{\bar{\bfp}},{\bar{\bfk}},\ell)\label{vsdw}\\
\mathcal{V}^{(\ell)}_{CDW}(\bfk,\bfp)&=&2g(\bfk,\bfp,{\bar{\bfp}},{\bar{\bfk}},\ell)-g(\bfp,\bfk,{\bar{\bfp}},{\bar{\bfk}},\ell)\nonumber\\
\mathcal{V}^{(\ell)}_{FM}(\bfk,\bfp)&=&-g(\bfp,\bfk,\bfp,\bfk,\ell)\nonumber\\
\mathcal{V}^{(\ell)}_{PI}(\bfk,\bfp)&=&2g(\bfk,\bfp,\bfp,\bfk,\ell)-g(\bfp,\bfk,\bfp,\bfk,\ell)  \  ,\nonumber
\end{eqnarray}
\noindent where $\bar{\bfk}={\bm k+\bm Q}$, $\bar{\bfp}={\bm p + Q}$, and ${\bm Q}=(\pi,\pi)$ is the nesting vector at half-filling, and the momenta, $\bfk$, $\bfp$, lie on either of the Fermi surfaces. 
The singlet(triplet) SC instability channel is denoted by $SC_s$($SC_{t_0}$) and corresponds to the plus(minus) sign in the right-hand side of the first equation above. 
In the numerical implementation, each Fermi surface is discretized into 36 patches and $\bfk$, $\bfp$ can be labeled by a discrete patch index that goes from 1 to $2\times 36$ [Fig.~\ref{fig:phase}(c)]. 
As defined above, the $\mathcal{V}^{(\ell)}_{\rm op}(\bfk,\bfp)$ contain both intra- and inter-species couplings, depending on whether the momenta $\bfk$ and $\bfp$ are on the same Fermi surface or on different Fermi surfaces, and thus $\bfk$ and $\bfp$ have 2$\times$36 values. 
Explicitly, each $\mathcal{V}$ matrix contains 4 blocks with species indices in the form 
$\left(\begin{matrix} g_{\rm ffff} & g_{\rm fccf} \cr g_{\rm cffc} & g_{\rm cccc} \end{matrix}\right)$
, and each block has dimension $36\times36$. 
For each order parameter channel, we can further diagonalize $\mathcal{V}^{(\ell)}_{\rm op}(\bfk,\bfp)$ into
\begin{equation}
\mathcal{V}^{(\ell)}_{\rm op}(\bfk,\bfp)=\sum_{i,oo^\prime}w_{\rm op}^{(i)}(\ell)\eta_o f^{(i)*}_{\rm op}(\bfk,o,\ell)\eta_{o^\prime} f^{(i)}_{\rm op}(\bfp,o^\prime,\ell), \nonumber
\end{equation}
with $i$ being a decomposition index, and where now we explicitly indicate the species index $o$ and $o^{\prime}$ for $c$- or $f$-fermions. 
The leading instability can be determined by the most negative eigenvalue $w_{\rm op}^{\rm min}$ (largest magnitude), and the corresponding symmetry of the instability ($s$-, $p$-, and $d$-wave etc.) is given by the form factor $f^{\rm min}_{\rm op}(\bfk)$. 
The sign structure factor $\eta_{o}=1(-1)$ stands for in-phase (out-phase) between species on the bipartite square lattice. 
For instance, the two $s$-wave form factors, $f^{\rm min}_{CDW}(\bfk,o)$ and $f^{\rm min}_{CDW}(\bfp,o^\prime)$, carry different signs, $\eta_o=1$ and $\eta_{o^\prime}=-1$, in the $s_{\pm}$-CDW phase in Fig.~\ref{fig:phase}(b).

\section{Results}\label{sct:result}
With fixed parameters (in units of $t$), $\mu_f=0$ (half-filling) and $U_{\rm ff}=U_{\rm cc}>0$, the phase diagram parameterized by $\mu_c$ and the dimensionless ratio of intra- and inter-species interaction $U_{\rm cf}/U_{\rm ff}$ is illustrated in Fig.~\ref{fig:phase}(b). 
For small inter-species interactions, the species at half-filling ($f$-fermions) has an SDW instability. With increasing $U_{\rm cf}/U_{\rm ff}$, if both species are at half-filling, an $s$-wave CDW emerges, where the order parameters for the two species have a $\pi$-phase difference, referred to as an $s_{\pm}$-CDW here. 
This is analogous to the $s_{\pm}$-wave pairing in iron-pnictide superconductors~\cite{FWang} but in the particle-hole channel. 
By relabeling  the square lattice as a bipartite lattice with A,B sub-lattices, the real-space picture of the $s_{\pm}$-CDW phase is that $f$($c$)-fermions only occupy $A(B)$ sub-lattice sites to avoid the strong inter-species repulsion.

Keeping the $f$-fermions at half-filling, we find phase transitions from $s_{\pm}$-CDW to $d_{xy}$-CDW by increasing $|\mu_c|$, and from SDW to $d_{xy}$-CDW by enhancing $U_{\rm cf}/U_{\rm ff}$. 
In Fig.~\ref{fig:phase}(c), we show the FS of the two species for $\mu_f=0$ and $\mu_c=-0.1$ and indicate the FS patches used in our numerical RG implementation. 
We also show, in Fig.~\ref{fig:phase}(d), the corresponding form factors of $d_{xy}$-CDW for $U_{\rm cf}/U_{\rm ff}=2.5$, obtained from decomposing the CDW coupling into eigenfunctions and plotting the dominant one close to the instability. 
The larger magnitude of the form factor for $f$-fermions indicates that the $d_{xy}$-instability is mainly driven by the species at half-filling, with ordering of the $c$-fermions resulting from a proximity effect. 
As we increase the density imbalance, the $d_{xy}$-CDW phase requires larger $U_{\rm cf}/U_{\rm ff}$ to dominate over SDW. The real-space picture of $d_{xy}$-CDW is sketched in Fig.~\ref{fig:phase}(a), where the crosses indicate higher densities along next-nearest-neighbor bonds of alternating plaquettes~\cite{Nayak}, 
$O_{\rm dCDW}(\bm{r})=e^{i\bm{Q}\cdot\bm{r}}\sum_{\alpha;i,j=\pm1}a^{\dag}_{\alpha}(x,y)a_{\alpha}(x+i,y+j)$. 

\begin{figure}[t]
\begin{center}
\includegraphics[width=7.2cm]{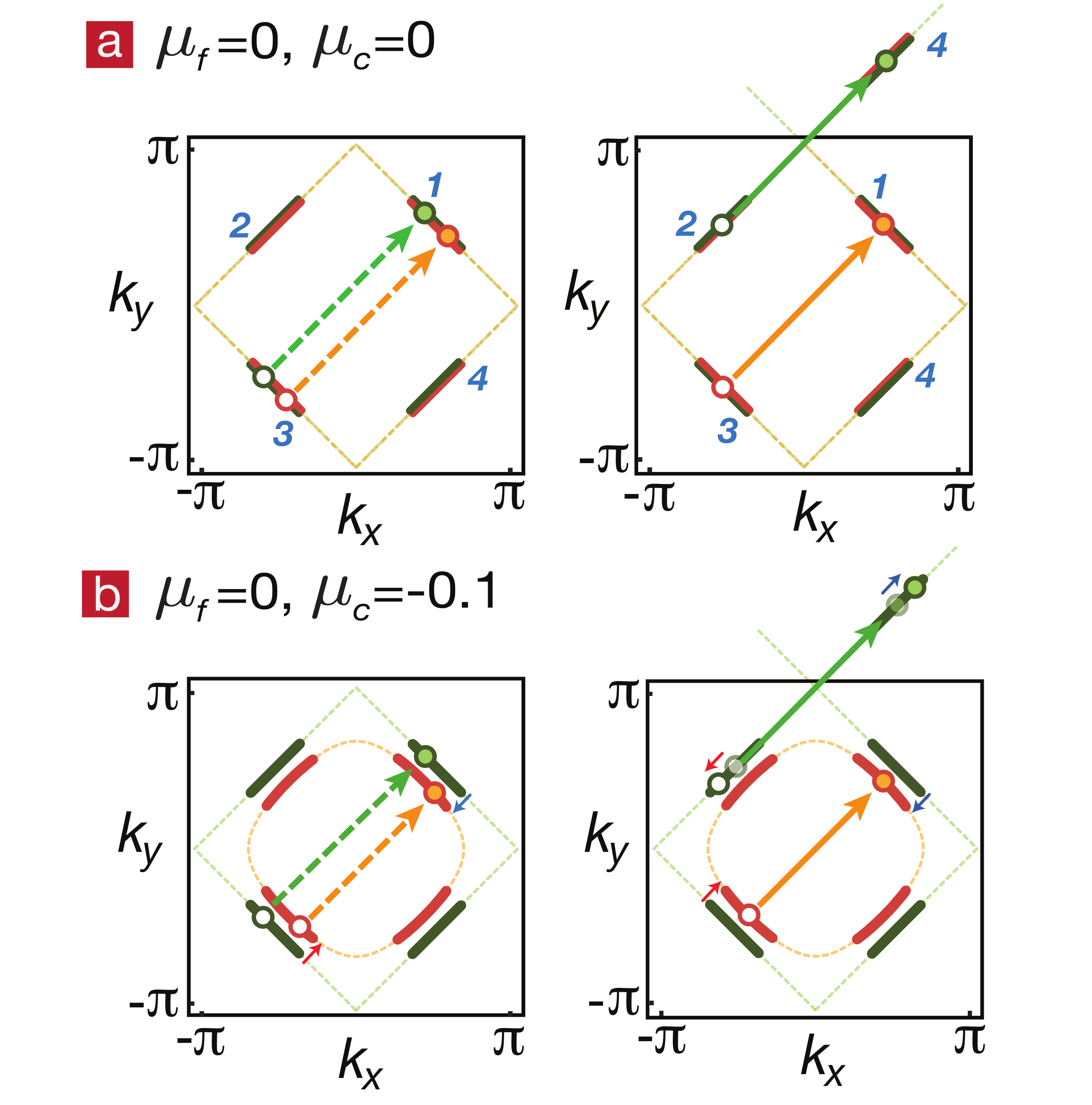}
\caption{(Color online) 
Sketches of two inter-species couplings, $g_1^{\rm cf}=g_{\rm cffc}(1,1,3,3)$ (left) and $g_{2}^{\rm cf}=g_{\rm cffc}(1,4,2,3)$ (right), for two different fillings of c-fermions. 
In (b), the deviation of the FS for $\mu_c=-0.1$ from the one at half-filling is intentionally enlarged for better visualization.}
\label{fig:gcf}
\end{center}
\end{figure}

\subsection{A Four-Patch g-ology Model}
To understand the physics leading to the development of the $d_{xy}$-CDW, we first explore the competition between the $s$-CDW and the $d_{xy}$-CDW of a single species ($f$-fermions) at half-filling. 
By decoupling the quartic interaction terms in the form $\sum_{\bm k,\bm p}\mathcal{V}^{(\ell)}_{\rm CDW}(\bm k,\bm p)\hat{\mathcal{O}}^{\dag}_{\bm k}\hat{\mathcal{O}}_{\bm p}$, with $\hat{\mathcal{O}}_{\bm k}$ being a term bilinear in the fermion fields and corresponding to the order parameter of CDW, we find that $\mathcal{V}_{\rm CDW}(\bm{k},\bm{p})\simeq g_{\rm ffff}(\bm{k},\bm{p},\bm{p}+\bm{Q},\bm{k}+\bm{Q})$ exhibits a negative eigenvalue when CDW dominates over the SDW~\cite{FWang}. 
In this case, for simplicity, we can divide the FS into four patches around $k_F=(\pi/2,\pi/2)$, $(-\pi/2,\pi/2)$ $(-\pi/2,-\pi/2)$ and $(\pi/2,-\pi/2)$ and consider only the scatterings occurring on these pieces of the FS [Fig.~\ref{fig:gcf}(a)]. 
Under the dihedral symmetry of a square lattice, three independent Umklapp couplings associated with momentum transfer $\bm{k}_4-\bm{k}_1=\bm{Q} = (\pi,\pi)$ can be defined: $g_1=g_{\rm ffff}(1,1,3,3)$, $g_{2}=g_{\rm ffff}(1,4,2,3)$ and $g_3=g_{\rm ffff}(1,3,1,3)$. 
The matrix of $\mathcal{V}_{\rm CDW}$ then takes the following form: 
\begin{eqnarray}
\mathcal{V}_{\rm CDW}\simeq\left(\begin{array}{cccc} 
g_{1} & g_{2} & g_3 & g_{2} \\
g_{2} & g_{1} & g_{2} & g_3 \\
g_3 & g_{2} & g_{1} & g_{2} \\
g_{2} & g_3 & g_{2} & g_{1}
\end{array}\right),
\end{eqnarray}
with eigenvectors $\bm{\Phi}_s=(1,1,1,1)$, $\bm{\Phi}_d=(1,-1,1,-1)$, $\bm{\Phi}_{p_1}=(1,0,-1,0)$ and $\bm{\Phi}_{p_2}=(0,1,0,-1)$, and corresponding eigenvalues $E_s=g_{1}+2g_{2}+g_3$, $E_d=g_{1}-2g_{2}+g_3$ and $E_{p_{1/2}}=g_{1}-g_3$. 
Thus, once the CDW instability is triggered, the competition between $d$- and $s$-wave channels is clear: $d_{xy}$-CDW ($s$-CDW) has the most negative eigenvalue and is therefore dominant when $g_{2}>0$ ($g_{2}<0$). 

\begin{figure}[!t]
\begin{center}
\includegraphics[width=6.7cm]{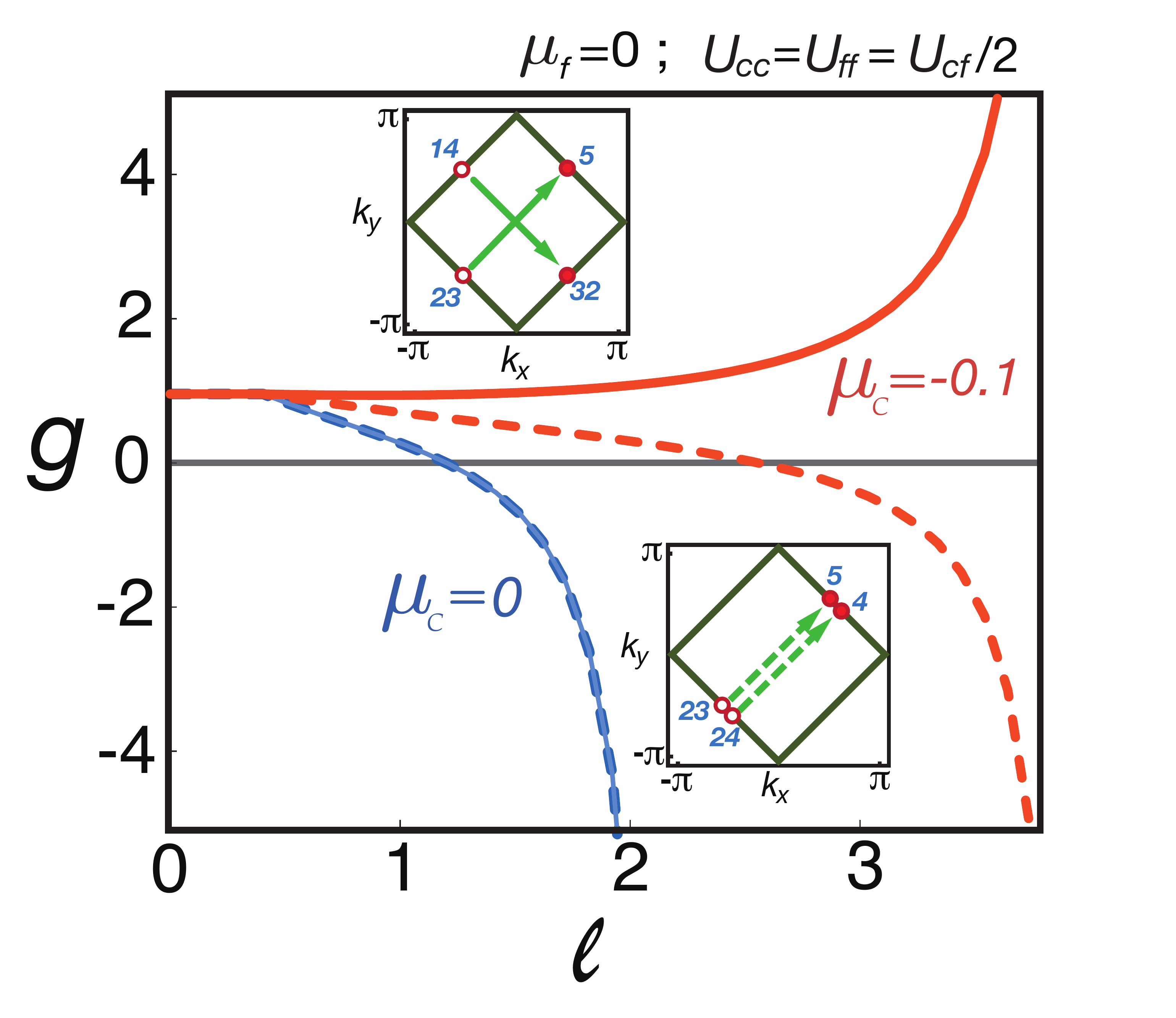}
\caption{(Color online) 
The flows of the most dominant couplings in the RG procedure, $g_{\rm ffff}(5,4,24,23)$ and $g_{\rm ffff}(5,32,14,23)$, are illustrated for $\mu_c=0$ (dashed and solid blue lines) -- note that these two lines are on top of each other, indicating that the two couplings have the same magnitude in this case; and $\mu_c=-0.1$ (dashed and solid orange lines). The two couplings are illustrated in the insets.}
\label{Umklapp}
\end{center}
\end{figure} 

We now introduce the second species of fermions ($c$-fermions), with filling slightly less than half and with the FS also divided into four patches (Fig.~\ref{fig:gcf}). 
By introducing an inter-species interaction and defining the Umklapp couplings $g_1^{\rm cf}=g_{\rm cffc}(1,1,3,3)$, $g_{2}^{\rm cf}=g_{\rm cffc}(1,4,2,3)$ and $g_3^{\rm cf}=g_{\rm cffc}(1,3,1,3)$, the one-loop RG equations for $g_1$ and $g_2$ 
are given by
\begin{eqnarray}
&&\hspace{-0.3cm}\frac{dg_{1}}{d\ell}=\mathcal{A}_{1jk}g_jg_k+\sum_{n=1}^3\mathcal{B}_{1nn}\left(g_n^{\rm cf}\right)^2+\mathcal{B}_{122}\left(g_2^{\rm cf}\right)^2,\label{eq:g1}\\
&&\hspace{-0.3cm}\frac{dg_2}{d\ell}=\mathcal{A}_{2jk}g_jg_k+2\mathcal{B}_{212}g_1^{\rm cf}g_2^{\rm cf}+2\mathcal{B}_{223}g_2^{\rm cf}g_3^{\rm cf}\label{eq:g2},
\end{eqnarray} 
where $\mathcal{A}_{ijk}$ and $\mathcal{B}_{ijk}$ are the kernels of the RG equations for intra- and inter-species couplings, respectively, and $i,j,k$ represent coupling indices. 
$\mathcal{A}_{ijk}$ contains contributions from all one-loop diagrams [Figs.~\ref{fig:rgeq}(b) to \ref{fig:rgeq}(f)]. 
$\mathcal{B}_{ijk}$ corresponds to a fermionic bubble [Fig.~\ref{fig:rgeq}(d)] for the $c$-fermions, obtained when two inter-species couplings are contracted to generate a correction to the intra-species $f$-fermion coupling. 
The $\mathcal{B}_{ijk}$'s are therefore always negative and correspond to an effective attractive interaction between the $f$-fermions, mediated by the $c$-fermions. 
Note that when the bare inter-species interaction is weak, $U_{\rm cf}/U_{\rm ff}\ll1$, and the $\mathcal{B}_{ijk}g_j^{\rm cf} g_k^{\rm cf} $ terms can be neglected. 
In this case, the $\mathcal{A}_{ijk}g_jg_k$ intra-species terms eventually drive $g_{1}$ and $g_{2}$ to large positive values, leading to an SDW instability.

\begin{figure*}[t!]
\begin{center}
\includegraphics[width=0.9\textwidth]{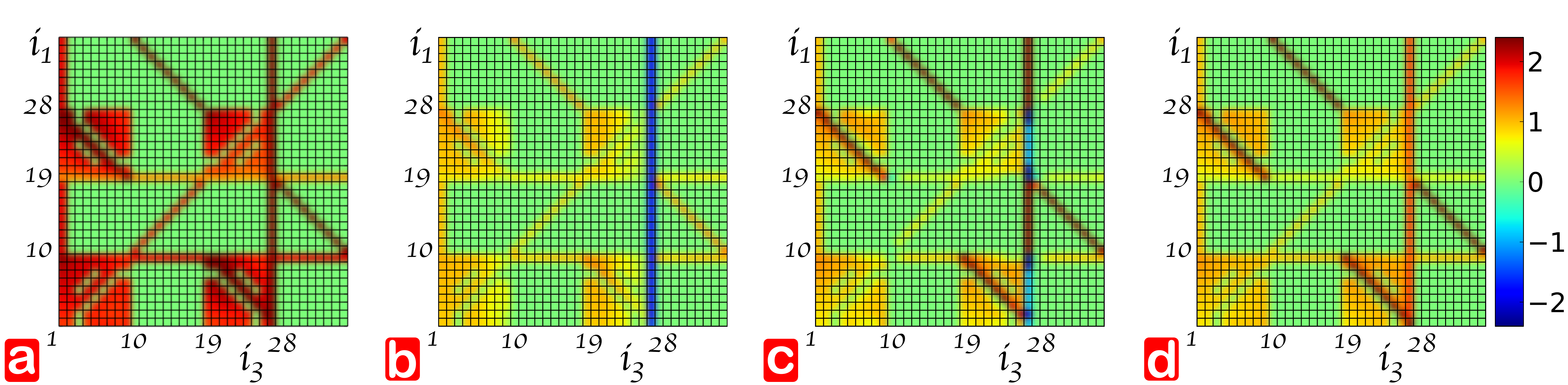}
\caption{(Color online) 
Snapshots of interaction vertices during RG flow for $f$-fermions at half-filling ($\mu_f=0$). 
(a) $g_{\rm cffc}(i_1,i_2=1,i_3)$ and (b)-(d) $g_{\rm ffff}(i_1,i_2=1,i_3)$. 
The other parameters are set to $\mu_c=0$, $U_{\rm cf}/U_{\rm ff}=2.0$ for (a) and (b), where the $s_{\pm}$-CDW instability occurs. 
In (c), $\mu_c=0.1t$ and $U_{\rm cf}/U_{\rm ff}=2.0$, this is in the regime where unconventional $d_{xy}$-wave CDW dominates. 
(d) $\mu_c=0.1t$, $U_{\rm cf}/U_{\rm ff}=0.5$, the SDW will be the instability if inter-species interaction is weak. 
Intra-species interactions are set to $U_{\rm cc}=U_{\rm ff}=1.0t$ in all cases shown here.}
\label{fig:vflow}
\end{center}
\end{figure*}

When both species are at half-filling, the two FS overlap and the three inter-species Umklapp couplings ($g_{1}^{\rm cf}$, $g_{2}^{\rm cf}$, $g_{3}^{\rm cf}$) exhibit equivalent phase space; the same applies to the intra-species Umklapp couplings. 
In this circumstance, increasing the bare value of the inter-species interaction in Eq.~(\ref{eq:g1}) and (\ref{eq:g2}) eventually leads both $g_{1}$ and $g_{2}$ to flow to negative values and gives rise the $s$-CDW instability. 
When one species is away from half-filling, the four patches on its rounded-square FS are slightly shifted from the ones at half-filling. 
In Fig.~\ref{fig:gcf}(b), we illustrate the two FS and corresponding momentum shifts between the patches. 
As shown in the left panel of Fig.~\ref{fig:gcf}(b), for fermions on the FS, the net momentum transfer in the inter-species process $g_{1}^{\rm cf}$ is no longer equal to a reciprocal lattice vector but has a small misfit corresponding to the shift in the position of the Fermi patches for the $c$-fermions. 
This is also true for the $g_{3}^{\rm cf}$ processes. 
As high-energy modes are eliminated during the RG transformation, these two couplings will have a reduced phase-space in which they can occur and are not allowed when only fermions on the FS remain. 
However, the misfit in the momentum transfer for scattering $c$-fermions across two parallel patches (momentum transfer ${\bm Q} - {\bm \delta}$) can be compensated in $g_{2}^{\rm cf}$ by having the $f$-fermions scatter by ${\bm Q} + {\bm \delta}$. 
As sketched in Fig.~\ref{fig:gcf}(b) (right panel), this scattering is allowed by momentum and energy conservation due to the nesting of the FS (flat parallel patches in our simple model). 
Thus, after moving one species away from half-filling, following Shankar's phase-space argument~\cite{Shankar:1994vy}, particle-hole diagrams involving $g_{2}^{\rm cf}$ have an extended phase space in which they are non-zero, while ones involving $g_{1}^{\rm cf}$ and $g_{3}^{\rm cf}$ can be ignored to leading order in $d\Lambda$. 
As a consequence, the RG equations, Eq.~(\ref{eq:g1}) and (\ref{eq:g2}), can be rewritten as
\begin{eqnarray}
  &&\frac{dg_1}{d\ell}=\mathcal{A}_{1jk}g_jg_k+2\mathcal{B}_{122}\left(g_2^{\rm cf}\right)^2\ , \\
  &&\frac{dg_2}{d\ell}=\mathcal{A}_{2jk}g_jg_k\ ,
\end{eqnarray}
where $g_2$ receives no contribution from inter-species terms. 
The inter-species interaction term still drives $g_1$ to a negative value (effective mediated attraction).
The other Umklapp interaction $g_2$, however, is renormalized only by intra-species repulsion and flows to positive values. 
The flow to negative values leads to CDW instability, and the asymmetry in the Umklapp components means that the CDW instability has $d_{xy}$-wave order parameter ($g_1<0$ and $g_2>0$).

\subsection{Vertex Flows under the full FRG}
Going back to the full FRG calculation with 2$\times$36 patches, we confirm the simple mechanism demonstrated with the four-patch model by identifying, among the RG flows for all the couplings, the ones that are most divergent and looking at their behavior.
In Fig.~\ref{Umklapp}, the RG flows of the most dominant intra-species couplings, $g_{\rm ffff}(5,4,24,23)$ and $g_{\rm ffff}(5,32,14,23)$, are plotted for two different cases: $c$-fermions at half-filling ($\mu_c=0$) and away from half-filling ($\mu_c = -0.1$). 
When both species are at half-filling, the two couplings are equal to each other and flow to large negative values (dashed and solid blue lines), leading to the $s$-CDW instability. 
However, as expected from the four-patch analysis, once the $c$-fermions are away from half-filling ($\mu_c=-0.1$), the two couplings are renormalized in opposite ways (dashed and solid orange lines), resulting in $d_{xy}$-CDW. 

Although we can explain the $d_{xy}$-CDW phase using a simple four-patch model and analytical arguments involving RG flows of just a few Umklapp couplings, it is instructive to see how some of the vertex functions renormalize.
In Fig.~\ref{fig:vflow}, we show a snapshot at a given ``RG-time" $\ell$ of a few coupling functions. 
Since each vertex $g_{abba}(i_1,i_2,i_3,\ell)$ is a function of three free patch-indices, we fix one of them, $i_2=1$, where the positions of each patch on the two Fermi surfaces are illustrated in Fig.~\ref{fig:phase}(c). 
For $f$-fermions at half-filling ($\mu_f=0$) and fixing $U_{\rm cc}=U_{\rm ff}= t$, we show couplings for different $\mu_c$ and $U_{\rm cf}/U_{\rm ff}$ corresponding to the $s_{\pm}$-CDW, $d_{xy}$-CDW, and SDW phases.
Fig.~\ref{fig:vflow}(a) and Fig.~\ref{fig:vflow}(b) show inter-species coupling $g_{\rm cffc}(i_1,i_2=1,i_3,\ell)$ and $f$-fermions intra-species coupling $g_{\rm ffff}(i_1,i_2=1,i_3,\ell)$, respectively, for $\mu_c=0$ and $U_{\rm cf}/U_{\rm ff}=2.0$, corresponding to a point in the phase diagram where $s_{\pm}$-CDW occurs [shown in Fig.~\ref{fig:phase}(a)]. 
Moving the $c$-fermions away from half-filling leads to a region of $d_{xy}$-CDW. 
Fig.~\ref{fig:vflow}(c) shows the intra-species coupling $g_{\rm ffff}(i_1,i_2=1,i_3,\ell)$ for this case ($\mu_c=0.1t$, $U_{\rm cf}/U_{\rm ff}=2.0$). 
Decreasing the inter-species coupling from this point so that $U_{\rm ff}>U_{\rm cf}$ leads to the SDW phase. 
The intra-species coupling $g_{\rm ffff}(i_1,i_2,i_3,\ell)$ for this case is shown in Fig.~\ref{fig:vflow}(d) ($\mu_c=0.1t$, $U_{\rm cf}/U_{\rm ff}=0.5$). 

It is easy to see from the panels in Fig.~\ref{fig:vflow} that flows to strong coupling under RG are related to Fermi surface nesting due to half-filling. 
There are two types of such couplings. 
The first type contains couplings of the form $g_{\rm ffff}(i_1,1,i_1+{\bf Q})$, which correspond to the four pieces of diagonal lines (dark red) in Fig.~\ref{fig:vflow}(d). 
These couplings account for the nesting between $\bfk_1$ and $\bfk_3$, and represent the SDW channel in Eq.~(\ref{vsdw}).
The second type contains $g_{\rm ffff}(i_1,1,27)$ couplings, which make up the vertical straight line at $i_3=27$ in all four panels of Fig.~\ref{fig:vflow} [blue line in Fig.~\ref{fig:vflow}(b), in particular], and are responsible for the CDW channel. 
One can also see some sub-dominant vertices in these figures. 
The pairing vertex, $g_{\rm ffff}(i_1,i_2=-i_1,i_3)$ corresponds to the horizontal line at $i_1=19$. 
The major diagonal line at 45$^\circ$, going from the lower left to the upper right corners of the panels, corresponds to the backward scattering, $g_{\rm ffff}(i_1,i_2,i_3=i_1)$. 
The forward scattering, $g_{\rm ffff}(i_1,i_2,i_3=i_2)$, is the vertical straight line located at $i_3=1$. 
Finally, there are four small squares, which are enhanced due to scattering of a particle from one patch to another belonging to the same flat side of the nested Fermi surface. 

\begin{figure}[!t]
\begin{center}
\includegraphics[width=0.45\textwidth]{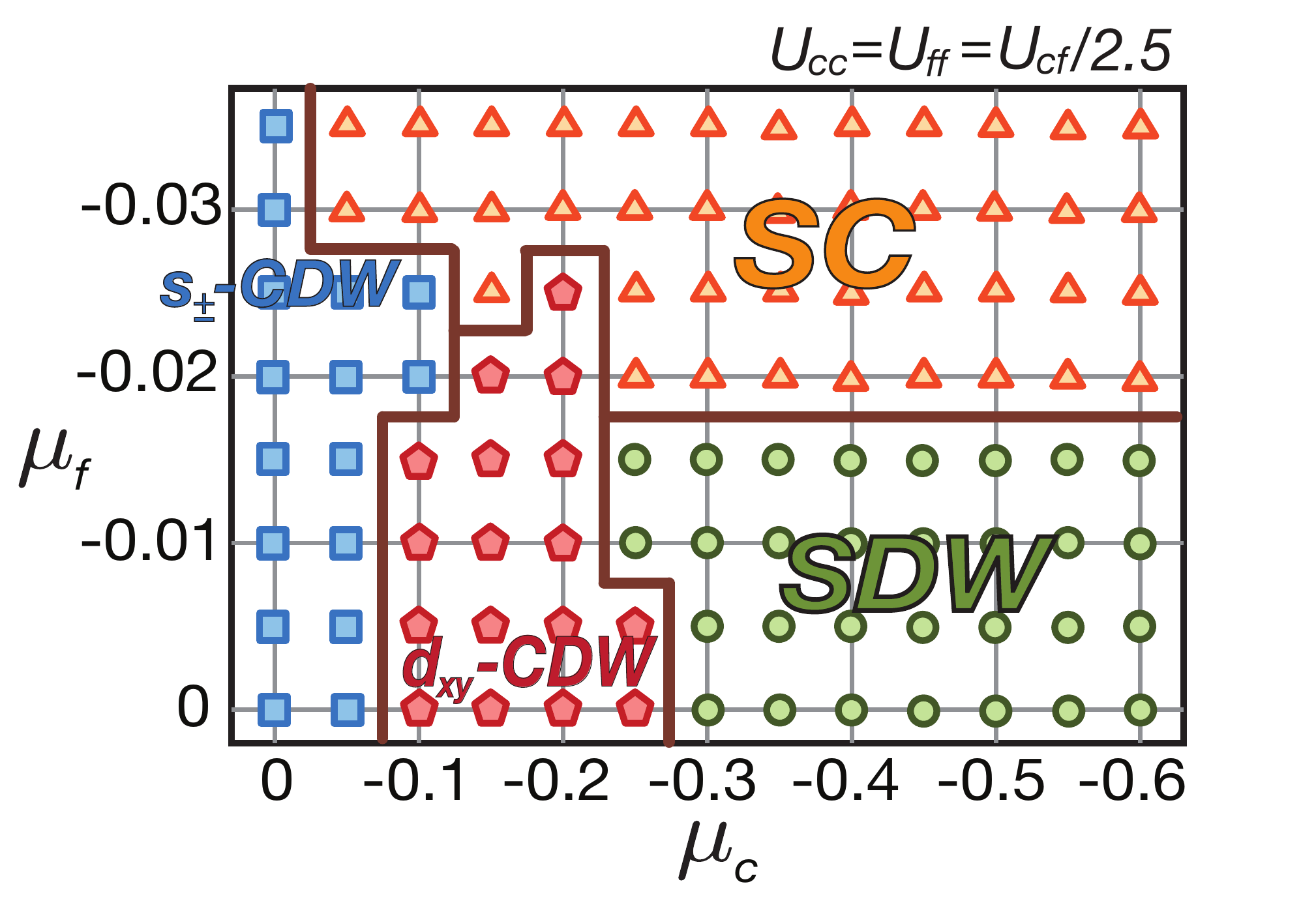}
\caption{(Color online) 
The phase diagram parameterized by $\mu_c$ and $\mu_f$. Note the large difference in scales between $\mu_c$ and $\mu_f$.}
\label{fig:PDmu}
\end{center}
\end{figure}

Starting from the $s_{\pm}$-CDW phase [Fig.~\ref{fig:vflow}(a) and \ref{fig:vflow}(b)], the leading divergent couplings are of the second kind, $g_{\rm ffff}(i_1,1,27)$, and flow to negative values [vertical line along $i_3=27$ in Fig.~\ref{fig:vflow}(b)], as we found in the four-patch model. 
The points along this vertical straight line of most divergent couplings have the same magnitude for all values of $i_1$, indicating an isotropic $s$-wave symmetry. 
Furthermore, the inter-species interaction, $g_{\rm cffc}(i_1,1,27)$, flows to strong positive values [Fig.~\ref{fig:vflow}(a)], again with the same value for all $i_1$ patch indices. 
This sign difference between $g_{\rm ffff}$ and $g_{\rm cffc}$ indicates that the $s$-wave order parameter has a $\pi$-phase between the $f$- and $c$-fermions, and thus this corresponds to an $s_{\pm}$-CDW instability. 

In Fig.~\ref{fig:vflow}(d), intra-species couplings are almost unaffected by weak inter-species couplings, and the strongest flow is in the SDW channel, which is represented by the four separate pieces of diagonal red lines. 
Finally, in Fig.~\ref{fig:vflow}(c), the couplings show that the nesting process, $g_{\rm ffff}(i_1,1,27)$, has alternating signs (red and blue in the vertical straight line at $i_3=27$). 
The sign alternation in the CDW form factor with four nodes, at $(\pm\pi,0)$ and $(0,\pm\pi)$, corresponds to $d_{xy}$-wave symmetry, instead of the more usual isotropic $s$-wave symmetry seen in Fig.~\ref{fig:vflow}(b). 
The physics behind the emergence of this sign change and $d_{xy}$-symmetry is as explained before. 
The population imbalance between the two fermion species in the vicinity of FS nesting provides the two FS with displaced parallel portions, which is the key ingredient for the mechanism we propose for creating a $d_{xy}$-CDW with purely repulsive interactions. 

The effective attractive interaction between $f$-fermions, mediated by the $c$-fermions, can also drive a BCS pairing instability if the nesting of the $f$-fermions FS is destroyed. 
To study the competition between pairing (SC) and density-wave instabilities, we explored the phase diagram parameterized by the chemical potentials of the two species, shown in Fig.~\ref{fig:PDmu} for $U_{\rm cc}=U_{\rm ff}=U_{\rm cf}/2.5$. 
The $d_{xy}$-CDW occurs as an intermediate phase between two limiting behaviors: $s_{\pm}$-CDW, when both species are at half-filling, and SDW, when one species is at half-filling and the other is far away from half-filling. 
Importantly, the $d_{xy}$ phase persists in a reasonably wide range of fillings of the minority species.

\subsection{The Critical Temperatures}
\begin{figure}
\begin{center}
\subfigure{\includegraphics[width=0.23\textwidth]{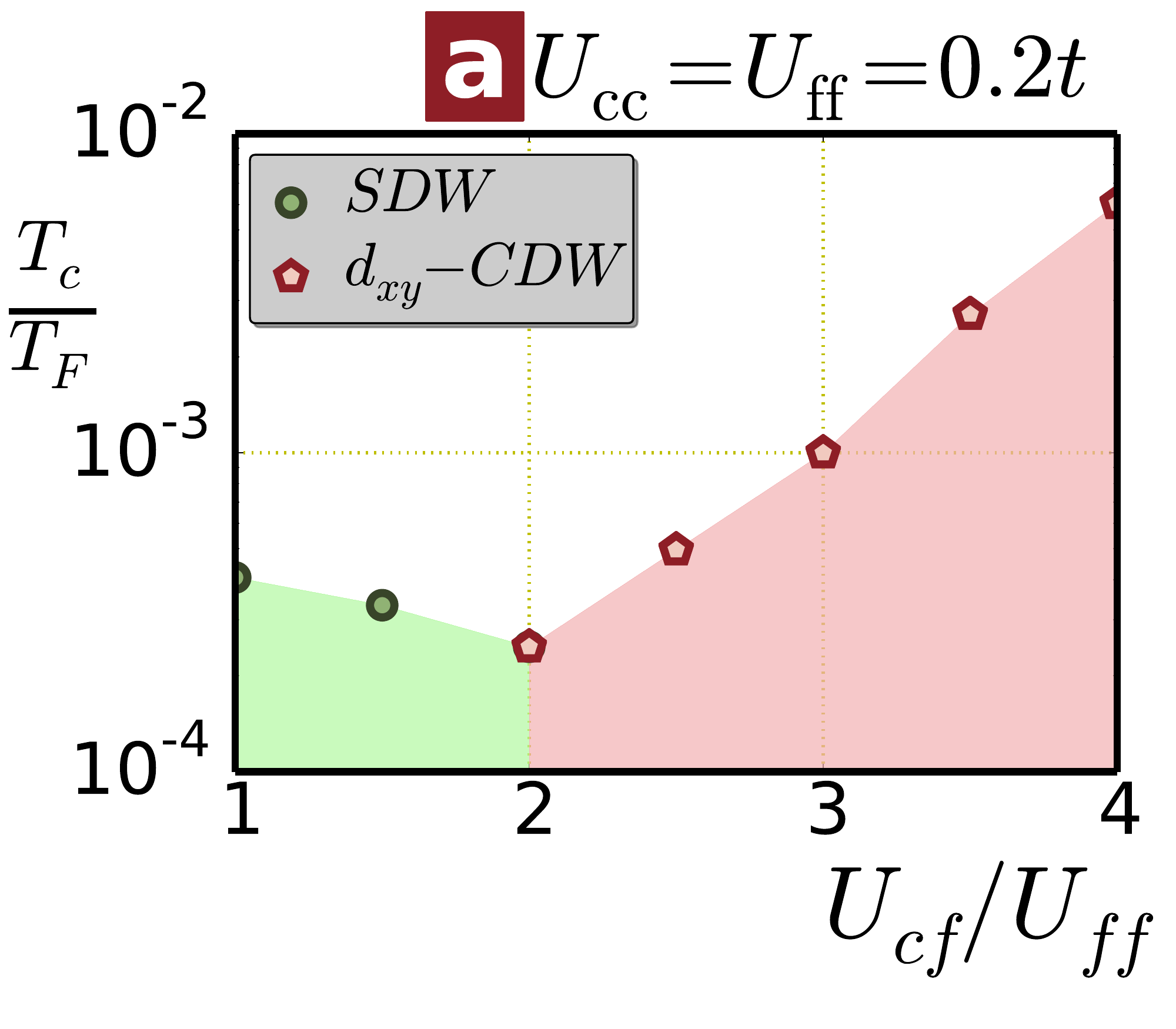}}
\subfigure{\includegraphics[width=0.23\textwidth]{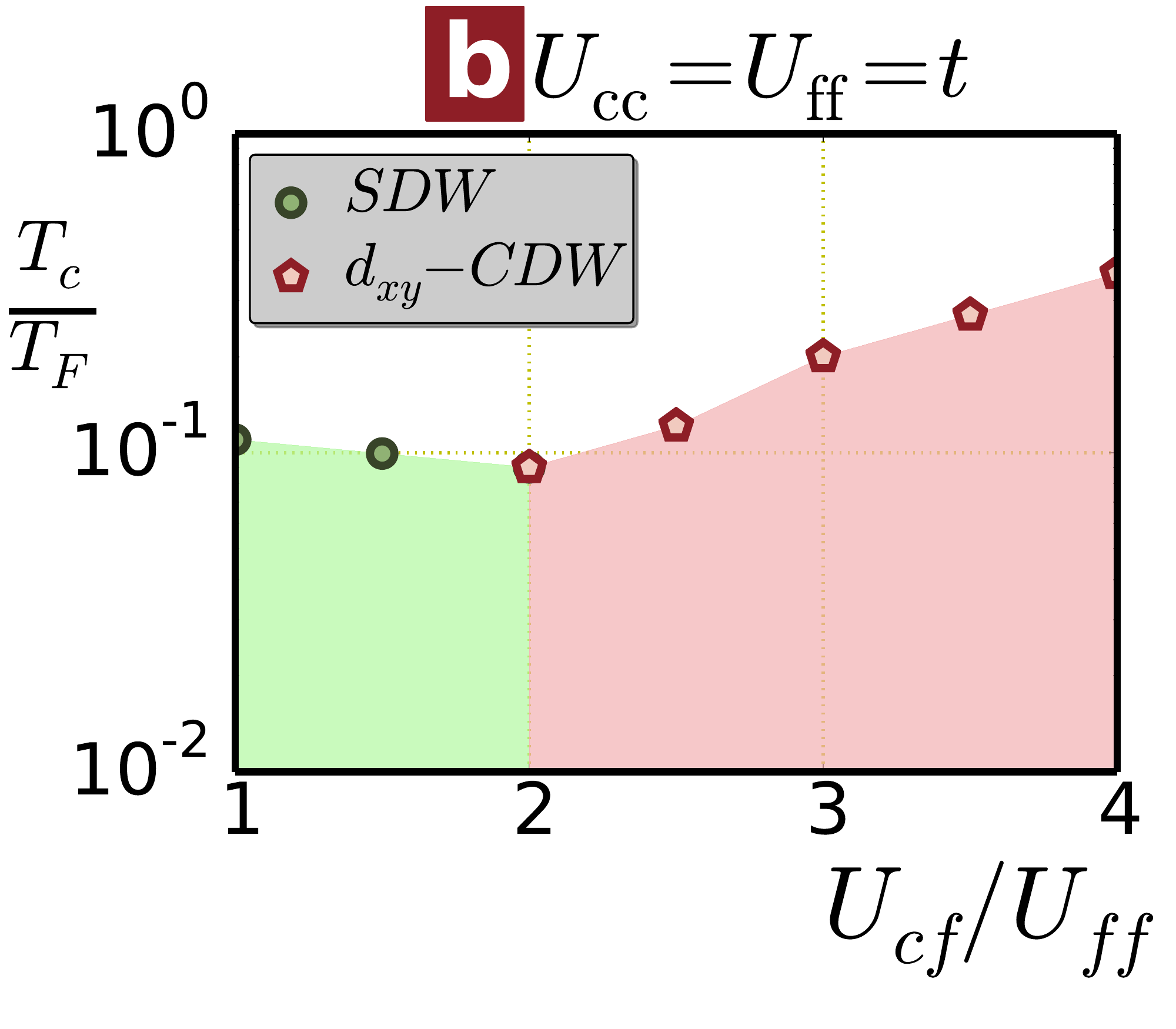}}
\caption{(Color online) 
Critical temperatures in units of the Fermi temperature $T_{\rm F}$ for the dominant instabilities SDW and $d_{xy}$-CDW as a function of the interaction ratio $U_{\rm cf}/U_{\rm ff}$, with $\mu_f=0$ and $\mu_c=0.2t$.}
\label{fig:tc}
\end{center}
\end{figure}
During the FRG calculation, the UV cutoff $\Lambda_l=\Lambda_0e^{-\ell}$ is reduced as the RG scale $\ell$ increases. 
The scale $\ell_c$ at which the divergence of the couplings occurs provides an estimate for the critical energy (and temperature) scale for the transition. 
We first use a small initial bare coupling, $U_{\rm cc}=U_{\rm ff}=0.2t$, and estimate the critical temperature for the $d_{xy}$-CDW transition. 
In Fig.~\ref{fig:tc}(a), we show the critical temperature from our FRG calculation as a function of the ratio $U_{\rm cf}/U_{\rm ff}$, for $U_{\rm cc}=U_{\rm ff}=0.2t$, for a fixed value of $\mu_c=0.2t$. 
This corresponds to a vertical cut of the phase diagram shown in ~\ref{fig:phase}(b). 
There are two competing orders, SDW and $d_{xy}$-CDW, and their corresponding $T_c$ can be given in units of the Fermi temperature $T_{\rm F}$. 
The critical temperatures found for this case are small, less than $10^{-2} T_{\rm F}$. 
However the critical temperature is significantly larger for larger bare interaction strengths, as shown in Fig.~\ref{fig:tc}(b) for $U_{\rm cc}=U_{\rm ff}=t$, where $T_c$ of the $d_{xy}$-CDW is more than 10\% of $T_{\rm F}$. 
Both figures [Fig.~\ref{fig:tc}(a) and (b)] show that the $T_c$ of SDW decreases as the ratio of initial bare interactions $U_{\rm cf}/U_{\rm ff}$ increases because it is stabilized by the on-site repulsion for the species at half-filling ($f$-fermions) and larger on-site inter-species repulsion $U_{\rm cf}$ acts against it. 
As the other competing order $d_{xy}$-CDW starts to dominate, its $T_c$ increases as the ratio increases. 
Larger initial on-site inter-species repulsion $U_{\rm cf}$ favors the $d_{xy}$-CDW instability. 
While the FRG approach only applies to weak-couplings, that is, bare coupling strengths small compared to the full energy bandwidth of the problem (of order $\mathcal{W}=8t$), the ratio of $U_{\rm cf}/U_{\rm ff}$ can be tuned to be very large, enhancing the value of $T_c$ for $d_{xy}$-CDW.

\section{Ultracold atom experiments}\label{sct:exp}
Our predicted phase diagram should be experimentally realizable with recently achieved quantum degenerate Li-K mixtures~\cite{Spiegelhalder:2010hq} or Li-Yb~\cite{Hara:2011gq} on a square optical lattice. 
We point out that a sufficiently large inter-species interaction $U_{\rm cf}$ compared to the intra-species repulsion of the species at half-filling $U_{\rm ff}$ is needed to create the $d_{xy}$-CDW we propose here. 
Taking the $^6$Li-$^{40}$K mixture, for example, with $^{40}$K as the $f$-fermions at half-filling, one can tune $U_{\rm ff}$ with a magnetic Feshbach resonance ($U_{\rm ff} \simeq 0$ at around 210 Gauss). 
At magnetic fields in this range, the $^6$Li - $^6$Li interaction $U_{\rm cc}$ is small and repulsive, as is the interspecies interaction $U_{\rm cf}$. 
One can therefore use the $^{40}$K Feshbach resonance to tune the $U_{\rm cf}/U_{\rm ff}$ ratio to large values, as required for our predicted $d_{xy}$-CDW phase to occur. 
One may also combine magnetic field with confinement-induced resonance~\cite{Petrov,Olshanii,Moritz}. 
To avoid breaking SU(2) symmetry, optical Feshbach resonances~\cite{Fedichev,Bohn} may also be explored. 
The all optical resonance has been used in fermionic systems~\cite{Blatt,Fu}, including the $^6$Li-$^{40}$K mixture~\cite{Spiegelhalder:2010hq}. 
Importantly, it is not necessary to have $U_{\rm ff} = U_{\rm cc}$ for our predictions to hold true. 
If these parameters are different, the phase diagram as a function of the ratio $U_{\rm cf}/U_{\rm ff}$ remains qualitatively unchanged, with no strong dependence on the value of $U_{\rm cc}$. 
Experimental candidates for realizing this mixture, such as $^6$Li-$^{40}$K mixture, have components with different values for their physical masses. 
The qualitative features of the phase diagram, however, do not require $t_{\rm f}=t_{\rm c}$. 
If the $c$-fermion, which is the species not at half-filling, is lighter than the $f$-fermion, the screening effect from $c$-fermions becomes even more effective than in the equal mass case, and the $d_{xy}$-CDW phase is further enhanced. 
If the $c$-fermions are heavier, the screening effect becomes less effective and, for $c$-fermions much heavier than $f$-fermions, the $d_{xy}$-CDW will eventually be completely suppressed. 
Therefore, for a $^6$Li-$^{40}$K mixture, having the $^{40}$K atoms be the $f$-fermions (half-filled) is more favorable for the emergence of $d_{xy}$-CDW order. 
These features provide more flexibility for the experimental realization of the $d_{xy}$-CDW phase we predict. 

Experiments with $^6$Li-$^{40}$K mixtures\cite{Taglieber} have achieved $T_{\rm Li} = 0.34 T_F$ and $T_{\rm K} = 0.40 T_F$. 
The values we find for $T_c$ using FRG are only rough estimates, but they are of the same order of magnitude as these experimental values. 
As the calculation of $T_c$ indicates, a sufficiently large inter-species interaction ($U_{\rm cf}$) compared to the intra-species repulsion of the species at half-filling ($U_{\rm ff}$) is needed to create the $d_{xy}$-CDW phase with a realistic value for $T_c$. 

Identification of order parameters with nonzero angular-momentum dependence is always a challenge topic in condensed matter systems, and requires measurement of the relative phase between different portions of the FS. 
%In ultracold atoms, s
Several methods have been proposed to perform such phase-sensitive measurements~\cite{Carusotto,Dao:2007gw,Gritsev,Pekker:2009vg,Kitagawa}, including a pump-probe scheme~\cite{Pekker:2009vg} and an approach based on noise correlations\cite{Kitagawa}. 
The $d_{xy}$-CDW order parameter symmetry may also be detected via momentum-resolved spectroscopy~\cite{Dao:2007gw,Stewart:2008kt}.

\section{Summary and Conclusion}\label{sct:conclusion}
In summary, we have studied a quantum degenerate Fermi-Fermi mixture on square lattice with different species density and interaction ratios. 
We have employed the  FRG, which goes beyond the mean-field approximation, to determine phase diagram of the system and to evaluate corresponding critical temperatures. 
By keeping one species at half-filling and varying the other, we found several phases, such as $s_\pm$-CDW, SDW, and most important $d_{xy}$-CDW. 
When inter-species density-density interaction is strong enough, system is dominated by charge density waves. 
For a small density imbalance between different species, our FRG result shows that the $d_{xy}$-CDW dominates over the conventional $s_\pm$-CDW. 
To clarify the physics behind the formation of this unconventional charge density wave, we considered an simple four-patch model, which shows clearly 
that the density imbalance makes Umklapp interactions flow asymmetrically. 
Along with FRG calculation, our study has allowed us to understand the structure of the phase diagram on a qualitative level. Finally, 
we have also proposed an experimental realization of our theoretical system involving  on Li - K mixture, and discussed how to fine tune the interaction ratios to achieve
the parameter range necessary to observe the phenomena we predict. Based on current cold atom technology, we believe this experiment can be realized in the near future.

\begin{acknowledgments}
We thank L. Mathey, E. Timmermans, and I. B. Spielman for fruitful discussions. 
We also thank P. Chen for computational resources from NSC and NCTS, Taiwan and the Triton Affiliates and Partners Program (TAPP) of the San Diego Supercomputer Center. 
CYL and SWT acknowledge support from NSF under Grant DMR-0847801, from the UC-Lab FRP under Award number 09-LR-05-118602, and from the Army Research Office, Department of Defense under Award Number W911NF-13-1-0119. 
WMH acknowledges support from NSC Taiwan under Grant 101-2917-I-564-074.
\end{acknowledgments}

\end{document}